\let\latexarabic\arabic
\let\latexdocument\document
\let\latexenddocument\enddocument

\RequirePackage[thmmarks]{ntheorem}
\makeatletter
\renewtheoremstyle{plain} 
  {\item[\hskip\labelsep \theorem@headerfont ##1\ \textup{##2}\theorem@separator]} 
  {\item[\hskip\labelsep \theorem@headerfont ##1\ \textup{##2}\ (##3)\theorem@separator]}
\makeatother

\documentclass[lineno]{biometrika}

\let\document\latexdocument
\let\enddocument\latexenddocument
\AtEndDocument{\printhistory}
\let\arabic\latexarabic
\def\rm{}

\usepackage{amsmath}
\usepackage{mathrsfs}  
\usepackage{amsfonts}
\usepackage{amssymb}
\usepackage{multirow}
\usepackage{graphicx}
\usepackage{makecell}
\usepackage{tabu}
\usepackage{threeparttable}
\usepackage{times}
\usepackage{bm}
\usepackage{natbib}
\usepackage{tabularx, booktabs} 
\usepackage[plain,noend]{algorithm2e}

\makeatletter
\renewcommand{\algocf@captiontext}[2]{#1\algocf@typo. \AlCapFnt{}#2} 
\def\@algocf@capt@plain{top}
\renewcommand{\algocf@makecaption}[2]{%
  \addtolength{\hsize}{\algomargin}%
  \sbox\@tempboxa{\algocf@captiontext{#1}{#2}}%
  \ifdim\wd\@tempboxa >\hsize
    \hskip .5\algomargin%
    \parbox[t]{\hsize}{\algocf@captiontext{#1}{#2}}
  \else%
    \global\@minipagefalse%
    \hbox to\hsize{\box\@tempboxa}
  \fi%
  \addtolength{\hsize}{-\algomargin}%
}
\makeatother

\newcommand\independent{\protect\mathpalette{\protect\independenT}{\perp}}
    \def\independenT#1#2{\mathrel{\rlap{$#1#2$}\mkern2mu{#1#2}}}

\begin{document}

\nolinenumbers


\markboth{Md. Abdul Basit, AHM Mahbub Latif \and Abdus S. Wahed}{Manuscript}

\title{A Risk-Ratio-Based Marginal Sensitivity Model for Causal Effects in Observational Studies}

\author{MD ABDUL BASIT, MAHBUB AHM LATIF}
\affil{Institute of Statistical Research and Training, \\ Univeristy of Dhaka, Bangladesh \email{abasit@isrt.ac.bd} \email{mlatif@isrt.ac.bd}}

\author{\and ABDUS S. WAHED}
\affil{Department of Biostatistics and Computational Biology, \\ University of Rochester, USA \email{abdus\_wahed@urmc.rochester.edu}}

\maketitle

\begin{abstract}

In observational studies, the identification of causal estimands depends on the no unmeasured confounding (NUC) assumption. As this assumption is not testable from observed data, sensitivity analysis plays an important role in observational studies to investigate the impact of unmeasured confounding on the causal conclusions. In this paper, we proposed a risk-ratio-based sensitivity analysis framework by introducing a modified marginal sensitivity model for observational studies with binary treatments. We further extended the proposed framework to the multivalued treatment setting. We then showed how the point estimate intervals and the corresponding percentile bootstrap confidence intervals can be constructed efficiently under the proposed framework. Simulation results suggested that the proposed framework of sensitivity analysis performs well in the presence of adequate overlap among the treatment groups. Lastly, we demonstrated our proposed sensitivity analysis framework by estimating the causal effect of maternal education on female fertility in Bangladesh.

\end{abstract}


\section{Introduction}

Randomized controlled trials (RCTs) are considered to be the gold standard for estimating the causal effect of a treatment on an outcome of interest. However, researchers are often limited to investigating causal relationships using only observational data when conducting randomized experiments is not feasible. In the absence of randomized treatment assignments, the estimation of causal effects in observational studies is usually based on a set of identifiability assumptions. One of the most important such assumptions is the strong ignorability or the no unmeasured confounding (NUC) assumption, which implies that we can observe all relevant confounders in an observational study. Since the NUC assumption is essentially unverifiable using observed data, it is necessary to perform sensitivity analyses that assess the robustness of the causal conclusions obtained from observational studies in the presence of unmeasured confounding. 

Sensitivity analysis is recognized as an essential step in the process of causal inference from observational studies. It refers to the approaches that investigate how causal conclusions are impacted in the presence of unmeasured confounding in observational studies. \citet{bind2021importance}, in their guidelines for the statistical reporting of observational studies, listed sensitivity analysis as one of the five major steps that need to be carried out before reporting causal conclusions elicited from observational studies. Many sponsors and regulatory bodies also nowadays require researchers to conduct some form of sensitivity analysis while drawing causal inferences from observational studies (for example, see standard MD-4 of the Patient-Centered Outcome Research Institute (PCORI) methodology standards for handling missing data at \texttt{https://tinyurl.com/kcbkvfwv}). 

\citet{cornfield1959smoking} conducted the first formal sensitivity analysis to investigate the impact of unmeasured confounding on the causal relationship between smoking and lung cancer. However, this initial sensitivity analysis framework was applicable to only binary outcomes, and it did not account for the sampling variation. Rosenbaum and colleagues overcame these limitations and greatly expanded the theory and methods for sensitivity analysis in a series of pioneering works \citep{rosenbaum1987sensitivity, gastwirth1998dual, rosenbaum2002covariance}. Recently, sensitivity analysis has gained a lot of research interest including \cite{bonvini2022sensitivity, dorn2021doubly, carnegie2016assessing, vanderweele2017sensitivity, zhao2019sensitivity, kallus2019interval, yadlowsky2022bounds}, among others. Most of these proposed sensitivity analysis frameworks deal with only binary treatments.

\citet{zhao2019sensitivity} recently proposed a sensitivity analysis framework for smooth estimators of causal effects, such as the inverse probability weighting (IPW) estimators. Their proposed marginal sensitivity model is a natural modification of Rosenbaum's sensitivity model \citep{rosenbaum2002b} for matched observational studies, and it quantifies the magnitude of unmeasured confounding by the odds ratio between the conditional probability of being treated given the measured confounders (observed propensity score) and conditional probability of being treated given both the measured and unmeasured confounders (true propensity score). It is well known that risk ratios are more consistent with the general intuition than odds ratios, and hence, are easier to interpret. Therefore, it is desirable to develop sensitivity analysis frameworks that interpret the sensitivity analysis results using risk ratios instead of odds ratios.

In this paper, we first modified the odds-ratio-based sensitivity analysis framework of \citet{zhao2019sensitivity} to a risk-ratio-based framework. In particular, we proposed a modified marginal sensitivity model that measures the violation of the NUC assumption using the risk ratio between the true propensity score and the observed propensity score (see Section \ref{sec:modified-sensitivity} for the exact definition). The proposed modified marginal sensitivity model introduces a new implicit sensitivity parameter that restricts the true propensity scores within its valid range for a given sensitivity model. We then extended the proposed modified marginal sensitivity model to the multivalued treatment setting based on the work of \citet{basit2023sensitivity}. We showed that the estimation of causal effects for binary and multivalued treatments can be performed efficiently under the proposed sensitivity analysis framework.

The rest of the paper is structured as follows: Section \ref{sec:zhao-sensitivity} describes necessary notations and briefly reviews the existing odds-ratio-based marginal sensitivity model; Section \ref{sec:rr-framework} introduces the proposed modified marginal sensitivity model for the binary treatment setting; Section \ref{sec:extension} extends the proposed sensitivity model to multivalued treatment settings; Section \ref{sec:multivalued_sim} reports results from simulation studies; Section \ref{sec:application} demonstrates sensitivity analysis under the proposed framework using a real data application; Section \ref{sec:conclusion} provides some concluding remarks. 



\section{The Sensitivity Analysis Framework}
\label{sec:zhao-sensitivity}

\subsection{The Potential Outcome Framework}

Consider an observational study with a binary treatment $A$ (1, if treated, and 0, if control), a vector of measured confounders $\pmb{X} \in \mathscr{X} \subset \mathbb{R}^{d}$, and the observed binary outcome {Y}, where $(A, \pmb{X}, Y) \sim F_0$. We observe $(A_{1}, \pmb{X}_{1}, Y_{1})$, $(A_{2}, \pmb{X}_{2}, Y_{2}), \ldots,(A_{n}, \pmb{X}_{n}, Y_{n})$, which denote data points observed from $n$ i.i.d. units from the true data generating distribution $F_0$. Moreover, let $Y_{i}(0)$ and $Y_{i}(1)$ denote the potential outcomes corresponding to treatment levels $0$ and $1$, respectively, for each unit $i \in [n]$. Under the stable unit treatment value assumption (SUTVA) \citep{rubin1978bayesian}, the observed outcomes can be defined in terms of potential outcomes as $Y_{i}=Y_{i}(A_{i})=A_{i} Y_{i}(1) + (1-A_{i}) Y_{i}(0)$. In order to estimate causal effects, we make the following identifiability assumptions:

\begin{assumption}[strong ignorability or NUC]
    There is no unmeasured confounding (NUC), i.e., $(Y(0), Y(1)) \independent A \,\vert\, \pmb{X}.$ In other words, the set of observed covariates, $\pmb{X}$ includes all common causes of $A$ and $Y$. 
    \label{assump: NUC}
\end{assumption}

\begin{assumption}[positivity or overlap]
    Each unit has a non-zero probability of receiving the treatment. That is, $0 < \mathbb{P}_0(A=1 \,\vert\, X=x) < 1, \thickspace \forall \thickspace \pmb{x} \in \pmb{X}$.
    \label{assump: overlap}
\end{assumption}

In observational studies with a binary treatment, one of the most commonly used causal estimands of interest is the average treatment effect (ATE), $\Delta:=\mathbb{E}_{0}[Y(1)]-\mathbb{E}_{0}[Y(0)]$, where $\mathbb{E}_0$ indicates that the expectation is taken over the true data generating distribution $F_0$.

If we define the observed propensity score as $e_{0}(\pmb{X}) = \mathbb{P}_0(A=1 \,\vert\, \pmb{X})$, then under Assumptions \ref{assump: NUC} and \ref{assump: overlap}, a consistent estimator of ATE $\Delta$ based on inverse probability weighting (IPW) is defined as 
\begin{align*}
    \hat{\Delta}_{\mathrm{IPW}}=\frac{1}{n} \sum_{i=1}^{n}
    \left[\frac{A_{i} Y_{i}}{\hat{e}\left(\pmb{X}_{i}\right)}-
    \frac{\left(1-A_{i}\right) Y_{i}}{1-\hat{e}\left(\pmb{X}_{i}\right)}\right],
\end{align*}
where $\hat{e}(\pmb{X})$ is a sample estimate of $e_{0}(\pmb{X})$. It is well known that the IPW estimates become unstable when the estimated propensity scores are close to $0$ or $1$ \citep{kang2007demystifying}. Therefore, the stabilized IPW (SIPW) estimator of ATE, obtained by multiplying the inverse probability weights by the probability of receiving the actual treatment, is frequently used in practice.

\subsection{The Odds-Ratio-Based Marginal Sensitivity Model}
\label{sec:marginal-model}

\citet{zhao2019sensitivity} developed a sensitivity analysis framework that can be used for smooth estimators of causal effects, e.g., the inverse probability weighting (IPW) estimators of ATE in observational studies with a binary treatment. Let us consider an unmeasured confounder $U$ that sums up all unmeasured confounding present in an observational study. \citet{robins2002covariance} suggested that the variable $U$ can be considered as any of the potential outcomes while defining the conditional probabilities of receiving treatments, i.e., propensity scores. So, let us denote the unobserved or true propensity score (conditional on both observed and unobserved confounders) by  $e_{a,0}(\pmb{x}, y)=\mathbb{P}_{0}\{A=a \,\vert\, \pmb{X}=\pmb{x}, Y(a)=y\}$ for $a \in \{0,1\}$ and the observed propensity score (conditional on observed confounders only) by $e_{a,0}(\pmb{x}) = \mathbb{P}_{0}\{A=a \,\vert\,\pmb{X}=\pmb{x}\}$. Then, the marginal sensitivity model assumes that
\begin{align}
    \frac{1}{\Lambda} \leqslant \operatorname{OR}\big\{e_{a}(\pmb{x}, y),
    e_{a,\pmb{\beta}_{0}}(\pmb{x})\big\} \leqslant \Lambda, \thickspace
    \forall \thickspace \pmb{x} \in \mathscr{X}, y \in \mathbb{R},
    a \in\{0,1\}, \label{eq:prIIobs}
\end{align}
where $\operatorname{OR}(p_1,p_2) = [p_1/(1-p_1)]/[p_2/(1-p_2)]$ is the odds ratio, $e_{a,\pmb{\beta}_{0}}(\pmb{x})$ is a parametric for $e_{a,0}(\pmb{x})$ for $a \in \{0,1\}$ using observed confounders, and $\Lambda \geqslant 1$ is a fixed sensitivity parameter that quantifies the magnitude of unmeasured confounding by the odds ratio of the true propensity score $e_a(\pmb{x},y)$ and the observed propensity score $e_{\beta_{a,0}}(\pmb{x})$ for $a \in \{0,1\}$. The marginal sensitivity model can also be defined nonparametrically by replacing $e_{a,\pmb{\beta}_{0}}(\pmb{x})$ with a nonparametric model $e_{a,0}(\pmb{x})$ in the sensitivity model defined in Expression \eqref{eq:prIIobs}.

Under the marginal sensitivity models, \citet{zhao2019sensitivity} used a generalized mini-max inequality and percentile bootstrap to convert the estimation problem of causal effects under any given sensitivity model to a linear programming problem (LPP), which can be solved very efficiently. 

\section{The proposed risk-ratio-based Framework of Sensitivity Analysis}
\label{sec:rr-framework}


\subsection{The Modified Marginal Sensitivity Model}
\label{sec:modified-sensitivity}
For observational studies with binary treatments, the marginal sensitivity model \eqref{eq:prIIobs} presented in Section \ref{sec:zhao-sensitivity} measures the violation of the NUC assumption in terms of the odds ratio between the observed and unobserved propensity score  $e_0(\pmb{x})$ and $e_0(\pmb{x}, y)$. Using the assumptions and notations defined in Section \ref{sec:zhao-sensitivity}, a natural modification of the marginal sensitivity model \eqref{eq:prIIobs} that quantifies the magnitude of the violation of the NUC assumption by the risk ratio between $e_{a,0}(\pmb{x})$ and $e_{a,0}(\pmb{x}, y)$ could be defined as $e_{a,0}(\pmb{x}, y) \in \mathcal{E}_{\beta_0}(\Gamma)$, where
\begin{align}
    \mathcal{E}_{\beta_0}(\Gamma)= \Big\{e_{a}(\pmb{x}, y):
    \frac{1}{\Gamma} \leq \operatorname{RR}\big\{e_{a,\beta_0}(\pmb{x}),
    e_a(\pmb{x}, y)\big\} \leq \Gamma, \forall
    \pmb{x} \in \mathscr{X}, y \in \mathbb{R}, a \in \{0,1\} \Big\},
    \label{eq:sens-model-mod}
\end{align}
where $\Gamma \geqslant 1$ is the sensitivity parameter and $\mathrm{RR}\left(p_{1}, p_{2}\right)= p_1 / p_2$ is the risk ratio.

Next, let us define
\begin{align*}
    k_{a,\beta_{0}}(\pmb{x})=\log \big(e_{a, \beta_{0}}(\pmb{x})\big), \;\;\; k_{a,0}(\pmb{x}, y)=\log \big(e_{a,0}(\pmb{x}, y)\big),   \;\;\; \text{ and } 
    \;\;\; k_{a}(\pmb{x}, y)=\log \big(e_{a}(\pmb{x}, y)\big),
\end{align*}
and let $l_{a,\beta_0}(\pmb{x}, y)=k_{a,\beta_0}(\pmb{x})-k_{a,0}(\pmb{x}, y)$ be the log-scale difference of the observed and the unobserved propensity scores. Similarly, for a postulated sensitivity model $e_a(\pmb{x},y)$, we define $l_a(\pmb{x}, y)=k_{a,\beta_0}(\pmb{x})-k_a(\pmb{x}, y)$. However, under the sensitivity model \eqref{eq:sens-model-mod}, the unobserved propensity score $e_a(\pmb{x}, y)$ does not lie within $(0, 1)$. Because, note that when $l_a(\pmb x, y) \rightarrow \infty$, then $\log \big(e_a(\pmb{x}, y)\big) \rightarrow -\infty$, and hence,  $e_a(\pmb{x}, y) \rightarrow 0$. But, as $l_a(\pmb x, y) \rightarrow -\infty$, then $\log \big(e_a(\pmb{x}, y)\big) \rightarrow \infty$, and hence,  $e_a(\pmb{x}, y) \rightarrow \infty$. That is, under the sensitivity model \eqref{eq:sens-model-mod}, the unobserved or true propensity score $e(\pmb{x}, y) \in (0, \infty)$.

To circumvent this problem, we introduce an additional sensitivity parameter in the sensitivity model \eqref{eq:sens-model-mod} that restricts $e_a(\pmb x, y)$ to be in its valid range $(0,1)$ as $l_a(x,y) \rightarrow -\infty$.

\begin{definition}[modified marginal sensitivity model]
\label{def:sens-model}
Fix a parameter $\Gamma_0 \geqslant 1$. Moreover, define $\Gamma_1 = \operatorname{max}\big\{e_{a,\beta_0}(\pmb{x}), \Gamma_{0}^{-1} \big\}$. In observational studies with binary treatments, the modified marginal sensitivity model assumes that $e_{a,0}(\pmb{x}, y) \in \mathcal{E}_{\beta_0}(\Gamma_0, \Gamma_1)$, where
\begin{align}
    \mathcal{E}_{\beta_0}(\Gamma_0, \Gamma_1) = \Big\{e_a(\pmb{x}, y)&: 
    \Gamma_1 \leq \operatorname{RR}\{e_{a,\beta_0}(\pmb{x}), 
    e_a(\pmb{x}, y)\} \leq \Gamma_0, \notag \\ &\;\;\;\;\;\;\;\;\forall \,
    \pmb{x} \in \mathscr{X}, y \in \mathbb{R}, a \in \{0,1\}\Big\},
    \label{eq:sens-model2}
\end{align}
and $\mathrm{RR}\left(p_{1}, p_{2}\right)= p_1 / p_2$ is the risk ratio.
\end{definition}

\begin{remark}
\label{remark: sens-parameter}
From the Definition \ref{def:sens-model}, we can see that the sensitivity parameter $\Gamma_1$ implicitly depends on the observed propensity score $e_{a,\beta_0}(\pmb{x})$ and the sensitivity parameter $\Gamma_0$. Therefore, we do not need to explicitly specify the value of $\Gamma_1$ while conducting a sensitivity analysis. Consequently, the introduction of the additional sensitivity parameter $\Gamma_1$ in the modified marginal sensitivity model does not necessarily increase the complexity of conducting the sensitivity analysis. The main purpose of the parameter $\Gamma_1$ is to restrict the unobserved propensity score $e_a(\pmb x, y)$ within its valid range $(0,1)$ under a specific sensitivity model.
\end{remark}

Now, Under the modified marginal sensitivity model \eqref{eq:sens-model2}, it is easy to observe that
$$
-\gamma_1 \leq l_a(\pmb{x},y) \leq \gamma_0,
$$
where $\gamma_0 = \log(\Gamma_0)$ and  $\gamma_1=\log(\Gamma_1) = \operatorname{max}\big\{\log(e_0(\pmb x)), -\gamma_0 \big\}$. 
Therefore, it can be shown that the sensitivity model \eqref{eq:sens-model2} is similar to assuming $l_a \in \mathcal{L}_{\beta_0}\left(\gamma_0, \gamma_1\right)$, where
$$
\mathcal{L}_{\beta_0}\left(\gamma_0, \gamma_1\right)=\left\{l_a: \mathscr{X} \times \mathbb{R} \rightarrow \mathbb{R} \text { and } \gamma_1 \leqslant l_a \leqslant \gamma_0, a \in \{0, 1\} \right\}.
$$
For the remainder of the paper, we consider $l_a(\pmb{x}, y)$ as sensitivity models for fixed sensitivity parameters $\gamma_0 \geqslant 0$.

\begin{remark}
    \label{remark:nonparametric}

We can also define the modified marginal sensitivity model \eqref{eq:sens-model-mod} nonparametrically by replacing $e_{a,\beta_0}(\pmb{x})$, and consequently, $k_{a,\beta_{0}}(\pmb{x})$ with corresponding nonparametric model $e_{a}(\pmb{x})$ and $k_{a}(\pmb{x})$, respectively. The statistical methods used in the proposed sensitivity analysis framework can be applied regardless of the choice of parametric or non-parametric model for $e_{a}(\pmb{x})$, as long as the model is smooth enough for a valid bootstrap.
\end{remark}


\subsection{Estimation of ATE under Modified Marginal Sensitivity models}

In order to estimate the ATE under the modified marginal sensitivity model, under a specific sensitivity model $l_a\in \mathcal{L}_{\beta_0}(\gamma_0, \gamma_1)$ for $a\in\{0,1\}$, let us define the shifted propensity scores as
\begin{align*}
    e^{(l_{a})}_a(\pmb{x}, y) &=
    \bigg[\exp \Big \{ l_{a}(\pmb{x}, y) - k_{a,\beta_0}(\pmb x) \Big \} \bigg]^{-1},
\end{align*}
and the shifted estimand of ATE as 
\begin{align}
    \Delta^{(l_0,l_1)}
    =&\Bigg\{\mathbb{E}_{0}\Bigg[\frac{A}
    {e^{(l_1)}_{1}(\pmb{X}, Y)}\Bigg]^{-1} \mathbb{E}_{0}\Bigg[\frac{A Y}
    {e^{(l_1)}_{1}(\pmb{X}, Y)}\Bigg]\Bigg\} \notag \\
    &-\Bigg\{\mathbb{E}_{0}\Bigg[\frac{1-A}
    {e^{(l_0)}_{0}(\pmb{X}, Y)}\Bigg]^{-1}  
    \mathbb{E}_{0}\Bigg[\frac{(1-A) Y}{e^{(l_0)}_{0}(\pmb{X}, Y)}\Bigg]\Bigg\}.
    \label{eq:shifted-ate}
\end{align}
Note that, for any $l_a\in \mathcal{L}_{\beta_0}(\gamma_0, \gamma_1)$, we have
\begin{align}
e^{(l_{a})}_a(\pmb{x}, y) &= \big[\exp \big(k_{a,\beta_0}(\pmb{x}) -k_a(\pmb x, y) - k_{a,\beta_0}(\pmb{x})\big)\big]^{-1} \notag \\
  &= \big[\exp \big(-\log (e_a(\pmb x, y)) \big) \big]^{-1} \notag \\
  &= \Bigg[ \frac{1}{e_a(\pmb x, y)} \Bigg] ^{-1} = e_a(\pmb x, y). \label{eq:shifted-ps}
\end{align}
That is, under any given sensitivity model $l_a$, our defined shifted propensity score is equivalent to the true propensity score $e_a(\pmb x, y)$. We can estimate these shifted propensity scores with
\begin{align*}
    \hat{e}^{(l_{a})}_a(\pmb{x}, y) &=
    \bigg[\exp \Big\{l_{a}(\pmb{x}, y) - \hat{k}_a(\pmb x) \Big\}\bigg]^{-1},
\end{align*}
where $\hat{k}_a(\pmb{x})=\hat{k}_{a,\hat{\pmb \beta}}(\pmb{x}) = \log \big(\hat{e}_{a, \beta}(\pmb x) \big)$ and $\hat{e}_{a, \beta}(\pmb x)$ is a parametric estimate of $e_{a, \beta}(\pmb x)$ for $a \in \{0, 1\}$.

Consequently, we can define the stabilized IPW (SIPW) estimate of $\Delta^{(l_0,l_1)}$ as
\begin{align} 
    \hat{\Delta}^{(l_0,l_1)}
    =&\Bigg\{\Bigg[\frac{1}{n} \sum_{i=1}^{n} 
    \frac{A_i}{\hat{e}^{(l_1)}_1\left(\pmb{X}_{i}, Y_{i}\right)}\Bigg]^{-1}\cdot
    \frac{1}{n} \sum_{i=1}^{n} 
    \frac{A_{i}Y_{i}}{\hat{e}^{(l_1)}_1\left(\pmb{X}_{i}, Y_{i}\right)}\Bigg\} \notag \\
    &- \Bigg\{\Bigg[\frac{1}{n} \sum_{i=1}^{n} 
    \frac{1-A_i}{\hat{e}^{(l_0)}_0\left(\pmb{X}_{i},
    Y_{i}\right)}\Bigg]^{-1}\cdot
    \frac{1}{n} \sum_{i=1}^{n} \frac{(1-A_i)Y_{i}}
    {\hat{e}^{(l_0)}_0\left(\pmb{X}_{i}, Y_{i}\right)}\Bigg\} \notag \\
    =&\frac{\sum_{i=1}^{n}A_{i}Y_{i}\:
    \hat{e}^{(l_1)}_1\left(\pmb{X}_{i}, Y_{i}\right)}
    {\sum_{i=1}^{n}A_{i}\:
    \hat{e}^{(l_1)}_1\left(\pmb{X}_{i}, Y_{i}\right)}
    -\frac{\sum_{i=1}^{n}(1-A_{i})Y_{i}\:
    \hat{e}^{(l_0)}_0\left(\pmb{X}_{i}, Y_{i}\right)}
    {\sum_{i=1}^{n}(1-A_{i})\:
    \hat{e}^{(l_0)}_0\left(\pmb{X}_{i}, Y_{i}\right)} \notag \\
    =&\frac{\sum_{i=1}^{n} A_{i}Y_{i}
    \big[\exp \{l_1(\pmb{X}_i,Y_i)-\hat{k}_1(\pmb{X}_{i})\}\big]}
    {\sum_{i=1}^{n} A_i\big[\exp \{l_1(\pmb{X}_i,Y_i)-
    \hat{k}_1(\pmb{X}_{i})\}\big]}  \notag \\
    &-\frac{\sum_{i=i}^{n} (1-A_{i})Y_{i}\big[
    \exp \{l_0(\pmb{X}_i,Y_i) - \hat{k}_0(\pmb{X}_{i})\}\big]}
    {\sum_{i=1}^{n}(1-A_i)\big[\exp \{l_0(\pmb{X}_i,Y_i)
    - \hat{k}_0(\pmb{X}_{i})\}\big]}. \label{eq:ate-hat2}
\end{align}

\citet{zhao2019sensitivity} showed that estimation problems such as the one given in Equation \eqref{eq:ate-hat2} can be transformed to a linear fractional programming (LFP) problem using Charnes-Cooper transformation \citep{charnes1962programming}. Before defining the LFP for Equation \eqref{eq:ate-hat2}, let us simplify the notations by assuming that the first $m \leqslant n$ units are in the treatment group $(A=1)$ and the rest of the observations are in the control group $(A=0)$, and that the outcomes are sorted in decreasing order among the first $m$ units and the other $n-m$ units. Finally, we can convert Equation \eqref{eq:ate-hat2} to the following LFP

\begin{equation}
\begin{aligned}
    \text{min or max} & \;\;\;\;\frac{\sum_{i=1}^{n} Y_{i}
    \big[z_{i} \exp \{-\hat{k}_1(\pmb{X}_{i})\}\big]}
    {\sum_{i=1}^{m}\big[z_{i} \exp \{-\hat{k}_1(\pmb{X}_{i})\}\big]}
    -\frac{\sum_{i=m+1}^{n} Y_{i}\big[z_{i}
    \exp \{-\hat{k}_0(\pmb{X}_{i})\}\big]}{\sum_{i=m+1}^{n}\big[z_{i}
    \exp \{-\hat{k}_0(\pmb{X}_{i})\}\big]} \\
    \text{subject to}&\;\;\;\; \Gamma_{1i} \leqslant z_{i} \leqslant \Gamma_0,
    \;\;\;\;\;\; \text { for } 1 \leqslant i \leqslant n, 
\end{aligned}
    \label{eq:lfp-ate2}
\end{equation}
where $z_{i}=\exp \big\{l_{1}\left(\pmb{X}_{i}, Y_{i}\right)\big\}$, for $1 \leqslant i \leqslant m,$ and $z_{i}=\exp \big\{l_{0}\left(\pmb{X}_{i}, Y_{i}\right)\big\},$ for $m+1 \leqslant i \leqslant n$. The LFP defined in Equation \eqref{eq:lfp-ate2} can further be transformed to a linear programming problem (LPP), which can be solved efficiently. The solution of the LPP \eqref{eq:lfp-ate2} yields a partially identified point estimate interval of ATE under a specific sensitivity model $l_a \in \mathcal{L}_{\beta_0}\left(\gamma_0, \gamma_1\right)$ and $a \in \{0, 1\}$. We can also obtain a $100(1 - \alpha)\%$ asymptotic confidence interval for ATE under a postulated sensitivity model $l_a$ using a percentile bootstrap approach \citep{zhao2019sensitivity}.


\section{Extension to Observational Studies with Multivalued Treatments}
\label{sec:extension}

In this section, we extend our proposed sensitivity analysis framework to the multivalued treatment setting. Suppose we observe i.i.d. $(A_{i}, \pmb{X}_{i}, Y_{i})_{i=1}^{n}$ from an observational study with $J > 2$ treatment levels, where
$A_{i} \in \mathcal{A} = \{1, 2, \ldots , J \}$ with the corresponding set of
potential outcomes $\mathcal{Y}_{i}=\{Y_{i}(1), Y_{i}(2), \ldots, Y_{i}(J)\}$, and
$\pmb{X}_{i} \in \mathscr{X} \subset \mathbb{R}^{d}$ is a vector of
observed confounders for each subject $i \in [n]$. Let us also define a treatment indicator $D_{i}(a)$ ($1$ if $A_{i}=a$, $0$ otherwise) for $a \in \mathcal{A}$. 

The identifiability assumptions for observational studies with multivalued treatments are almost equivalent to those in the binary treatment setting. We assume the overlap assumption that implies $\mathbb{P}_{0}(A=a \vert \pmb{X}=\pmb{x})>0$ for all $a \in \mathcal{A}$. However, instead of the strong ignorability assumption, we assume weak \textit{ignorability} \citep{imbens2000role}.

\begin{assumption}[weak ignorability]
    $D(a) \independent Y(a) \thinspace \vert \thinspace \pmb{X}, \thickspace \forall \thinspace a \in \mathcal{A}.$
\end{assumption}

\citet{imbens2000role} extended the concept of propensity scores from binary treatments to multivalued treatments introducing the generalized propensity score (GPS), which is defined as $r_{a,0}(\pmb{x}) = \mathbb{P}_{0}(A=a \vert \pmb{X}=\pmb{x})$ for $a \in \mathcal{A}$. As in Section \ref{sec:rr-framework}, let us further denote the unobserved or true propensity score as $r_{a,0}(\pmb{x}, y) = \mathbb{P}_{0}(A=a \vert \pmb{X}=\pmb{x}, Y(a) = y)$.

Based on these assumptions and notations, \citet{basit2023sensitivity} recently proposed a sensitivity analysis framework for the multivalued treatment setting extending the framework of \citet{zhao2019sensitivity} for binary treatments. Using similar ideas, we propose the following risk-ratio-based modified marginal sensitive model for observational studies with multivalued treatments.
\begin{definition}[modified marginal sensitivity model for multivalued treatments]
\label{def:sens-model2}
For fixed $\Gamma_0 \geqslant 1$ and $\Gamma_1 = \operatorname{max}\big\{r_{a,\beta_0}(\pmb{x}), \Gamma_{0}^{-1} \big\}$, the modified marginal sensitivity model for multivalued treatments assumes that 
\begin{align}
    \Gamma_1 \leqslant \mathrm{RR}\big\{r_{a,\beta_0}(\pmb{x}),
    r_a(\pmb{x}, y) \big\} \leqslant \Gamma_0, \thickspace \forall \thickspace 
    \pmb{x} \in \mathscr{X},a \in \mathcal{A},y \in \mathbb{R},
    \label{eq:multi_parasensem2}
\end{align}

where $\Gamma_0$ and $\Gamma_1$ are sensitivity parameters and $\mathrm{RR}\left(p_{1}, p_{2}\right)= p_1 / p_2$ is the risk ratio.
\end{definition}

Next, as in section \ref{sec:modified-sensitivity}, let us define
\begin{align*}
    k_{a,\beta_0}(\pmb{x}) = \log
    \big(r_{a,\beta_0}(\pmb{x})\big) \;\;\; \text { and }
    \;\;\; k_a(\pmb{x}, y) = \log \big(r_a(\pmb{x}, y)\big),
\end{align*}
and let $l_a(\pmb{x}, y) = k_{a,\beta_0}(\pmb{x}) - l_a(\pmb{x}, y)$ for all $a \in \mathcal{A}$ be the log-scale difference between the observed and the unobserved GPSs under any specified sensitivity model $r_a(\pmb{x}, y)$. Then, it can be shown that the sensitivity model \eqref{eq:multi_parasensem2} is equivalent to assuming $l_a \in \mathcal{L}_{\beta_0}(\gamma_0, \gamma_1)$, where
\begin{align*}
    \mathcal{L}_{\beta_0}(\gamma_0, \gamma_1)&=\big\{l_a: \mathscr{X} \times \mathbb{R}
    \rightarrow \mathbb{R} \text { and } \gamma_1 \leqslant l_a \leqslant
    \gamma_0,\;a\in \mathcal{A}\big\},
\end{align*}
$\gamma_0 = \log (\Gamma_0)$, and $\gamma_1=\log(\Gamma_1) = \operatorname{max}\big(\log(r_{a,\beta_0}(\pmb{x})), -\gamma_0 \big)$ are the sensitivity parameters.

In order to define estimands of causal effects in the multivalued treatment setting, we use a general class of additive causal estimands proposed by \citet{basit2023sensitivity}. This class of estimands is based on inverse probability weighting and is defined as
\begin{equation}
  \tau(\pmb{c}) = \sum_{a = 1}^{J} c_a m(a) = \sum_{a = 1}^{J} c_a \mathbb{E}_0\Bigg[\frac{Y \thinspace D(a)}{r_a(\pmb{X})}\Bigg],
  \label{eq:contrasts}
\end{equation}
where $\pmb{c}=(c_1, \ldots, c_J)'$ is a vector of contrasts,  $m(a) = \mathbb{E}_0[Y(a)]$ is the average potential outcome for $a \in \mathcal{A}$. These estimands encompass many commonly used causal estimands of interest for multivalued treatments, such as the pairwise average treatment effects (PATEs). 

Based on the identifiability assumptions defined for multivalued treatments, we define the shifted estimand of the average potential outcome $m(a)$ under a specified sensitivity model $l_a$ as
\begin{align}
m^{(h_a)}(a) = \mathbb{E}_0\Bigg[\frac{Y D(a)}{r_a(\pmb{X})}\Bigg]
= \mathbb{E}_0\Bigg[\frac{Y D(a)}{r_a(\pmb{X}, Y)}\Bigg]
= \mathbb{E}_0\Bigg[\frac{Y D(a)}{r^{(h_a)}(\pmb{X}, Y)}\Bigg],
\label{shifted_estimator}
\end{align}
where $r^{(l_a)}(\pmb{x}, y)=\big[\exp \big(l_a(\pmb{x}, y)-k_{a,\beta_0}(\pmb{x})\big)\big]^{-1}$ is the shifted GPS.
Consequently,  under the modified marginal sensitivity model \eqref{eq:multi_parasensem2}, the shifted causal estimand becomes
\begin{align}
  \tau^{(l_a)}(\pmb{c}) = \sum_{a = 1}^{J} c_a m^{(l_a)}(a).
  \label{eq:shifted_contrasts2}
\end{align}

Similar to the estimation of the shifted ATE defined in Equation \eqref{eq:shifted-ate} for binary treatments, we can show that the estimation of the causal estimand can be converted to a linear programming problem (LPP) that allows us to efficiently estimate the partially identified point estimate intervals of $\tau(\pmb{c})$ for any $l_a \in \mathcal{L}_{\beta_0}(\gamma_0, \gamma_1)$. The percentile bootstrap approach of \cite{zhao2019sensitivity} is also applicable for the computation of $100(1 - \alpha)\%$ asymptotic confidence intervals for $\tau(\pmb{c})$ \citep{basit2023sensitivity}.

\section{Simulation Study}
\label{sec:multivalued_sim}

We conducted simulated studies to investigate the performance of the proposed sensitivity analysis framework in the multivalued treatment setting. Our data generating mechanism and simulation settings are equivalent to \citet{basit2023sensitivity}. We simulate three covariates as $X_{i1} \sim \operatorname{Bernoulli}(0.5)$,
$X_{i2} \sim \operatorname{U}(-1, 1)$, and $X_{i3} \sim \operatorname{N}(0, 0.5)$. For each $i \in[n]$,
the covariate vector then becomes $\pmb{X}_{i} = (1, X_{i1}, X_{i2}, X_{i3}) ^ {T}$.
We simulated the treatment assignment mechanism using the following multinomial distribution
$$
\big(D_{i}(1), D_{i}(2), D_{i}(3)\big) \;\big\vert\; \pmb{X}_{i} \sim 
\operatorname{Multinom}\big(r_{1}(\pmb{X}_{i}), r_{2}(\pmb{X}_{i}), r_{3}(\pmb{X}_{i})\big),
$$
where $D_{i}(a)$ is the treatment indicator and
$$
r_{a}(\pmb{X}_{i}, Y_i) = r_{a}(\pmb{X}_{i}) = \frac{\exp \big(\pmb{X}_{i}^{T} \beta_{a}\big)}{\sum_{a^\prime=1}^{3} \exp \big(\pmb{X}_{i}^{T} \beta_{a ^ \prime}\big)} 
$$
denotes the complete or unobserved GPSs for $a \in \{1,2,3\}$ with $\beta_1 = (0, 0, 0, 0)^T$, $\beta_2 = k_2 \times (0, 1, 1, 1)^T$,
and $\beta_3 = k_3 \times (0, 1, 1, -1)^T$. In order to assess the influence of the degree of overlap on the point estimates and confidence intervals of our proposed frameworks, we considered
two simulation scenarios. In the first scenario,
we set $(k_2, k_3) = (0.1, -0.1)$ to simulate a scenario with adequate overlap in
the covariates, and in the second scenario, we set $(k_2, k_3) = (3, 3)$ to induce lack of overlap.
The potential outcomes are generated from the following multinomial distribution

$$
\big(Y_{i}(1), Y_{i}(2), Y_{i}(3)\big) \;\big\vert\; \pmb{X}_{i} \sim 
\operatorname{Multinom}\big(p_{Y_1}(\pmb{X}_{i}), p_{Y_2}(\pmb{X}_{i}), p_{Y_3}(\pmb{X}_{i})\big),
$$
where $Y_i(a)$ is the potential outcome for treatment level $a$ and 

$$
p_{Y_a}(\pmb{X}_{i}) = \mathbb{P}(Y(a) = 1 \big \vert \pmb{X}_{i}) = \frac{\exp \big(\pmb{X}_{i}^{T} \delta_{a}\big)}{\sum_{a^\prime=1}^{3} \exp \big( \pmb{X}_{i}^{T} \delta_{a^\prime}\big)} 
$$
with $\delta_1 = (1, 1, 1, 1)^T$,
$\delta_2 = (1, 1, -1, 1)^T$, and $\delta_3 = (1, 1, 1, -1)^T$. The observed outcome was then obtained as $Y_i = \sum_{a=1}^{J} D_i(a) Y_i(a)$ for each subject $i \in [n]$. We simulated $1000$ datasets with sample size $n=750$ for each scenario and estimate the interval of point
estimates and confidence intervals for the pairwise ATEs using our proposed sensitivity analysis framework. The pairwise ATEs are denoted by $\tau_{i,j}$, for
${i,j} \in \{1,2,3\}$. We considered six values of the sensitivity parameters $\gamma_0$, namely, $\gamma_0 = \{0, 0.1, 0.2, 0.5, 1, 2\}$.
The true partially identified intervals were obtained under each simulation scenario using large scale numerical approximations.
Since the treatment under consideration is multivalued with three treatment levels, the observed generalized
propensity scores (GPSs) are modeled using the multinomial logit regression model.

\begin{table}[!h]

\caption{Simulation results for the sensitivity analysis of pairwise ATEs under the RR framework in an observational study with three treatment levels}
\label{tab:multivalued-RR}
\centering
\resizebox{\linewidth}{!}{
\begin{threeparttable}
\begin{tabular}[t]{ccrrrrrrrr}
\toprule
\multicolumn{4}{c}{ } & \multicolumn{2}{c}{\% Bias} & \multicolumn{4}{c}{ } \\
\cmidrule(l{3pt}r{3pt}){5-6}
Scenario & ATE & $\gamma_0$ & $\Gamma_0$ & Lower & Upper & \makecell[c]{Non-\\ coverage} & \makecell[c]{Partially identified \\ interval} & \makecell[c]{Point estimate \\ interval} & \makecell[c]{Confidence \\ interval}\\
\midrule
 &  & 0.0 & 1.00 & $2.79$ & $2.79$ & 0.103 & $(-0.050, -0.050)$ & $(-0.048, -0.048)$ & $(-0.115, 0.017)$\\

 &  & 0.1 & 1.11 & $1.76$ & $2.88$ & 0.099 & $(-0.137, 0.038)$ & $(-0.136, 0.040)$ & $(-0.201, 0.106)$\\

 &  & 0.2 & 1.22 & $2.64$ & $1.72$ & 0.108 & $(-0.223, 0.126)$ & $(-0.223, 0.127)$ & $(-0.285, 0.193)$\\

 &  & 0.5 & 1.65 & $4.79$ & $0.76$ & 0.116 & $(-0.460, 0.375)$ & $(-0.459, 0.376)$ & $(-0.512, 0.436)$\\

 &  & 1.0 & 2.72 & $11.14$ & $-6.18$ & 0.123 & $(-0.744, 0.694)$ & $(-0.741, 0.693)$ & $(-0.771, 0.729)$\\

 & \multirow{-6}{*}{\centering\arraybackslash $\tau_{1, 2}$} & 2.0 & 7.39 & $5.03$ & $-7.64$ & 0.095 & $(-0.900, 0.882)$ & $(-0.900, 0.882)$ & $(-0.915, 0.900)$\\
\cmidrule{2-10}
 &  & 0.0 & 1.00 & $1.49$ & $1.49$ & 0.108 & $(-0.034, -0.034)$ & $(-0.035, -0.035)$ & $(-0.102, 0.032)$\\

 &  & 0.1 & 1.11 & $2.56$ & $1.42$ & 0.107 & $(-0.121, 0.053)$ & $(-0.121, 0.053)$ & $(-0.186, 0.119)$\\

 &  & 0.2 & 1.22 & $4.83$ & $2.18$ & 0.109 & $(-0.207, 0.139)$ & $(-0.207, 0.140)$ & $(-0.271, 0.205)$\\

 &  & 0.5 & 1.65 & $6.45$ & $-1.08$ & 0.113 & $(-0.444, 0.385)$ & $(-0.443, 0.385)$ & $(-0.499, 0.444)$\\

 &  & 1.0 & 2.72 & $9.69$ & $-5.88$ & 0.117 & $(-0.735, 0.699)$ & $(-0.733, 0.697)$ & $(-0.766, 0.734)$\\

 & \multirow{-6}{*}{\centering\arraybackslash $\tau_{1, 3}$} & 2.0 & 7.39 & $10.19$ & $-1.84$ & 0.102 & $(-0.903, 0.885)$ & $(-0.903, 0.885)$ & $(-0.918, 0.904)$\\
\cmidrule{2-10}
 &  & 0.0 & 1.00 & $1.22$ & $1.22$ & 0.097 & $(0.015, 0.015)$ & $(0.016, 0.016)$ & $(-0.052, 0.084)$\\

 &  & 0.1 & 1.11 & $0.78$ & $-1.32$ & 0.098 & $(-0.075, 0.106)$ & $(-0.074, 0.106)$ & $(-0.142, 0.173)$\\

 &  & 0.2 & 1.22 & $2.95$ & $-0.14$ & 0.105 & $(-0.165, 0.194)$ & $(-0.164, 0.195)$ & $(-0.230, 0.260)$\\

 &  & 0.5 & 1.65 & $2.88$ & $-2.97$ & 0.111 & $(-0.414, 0.440)$ & $(-0.413, 0.440)$ & $(-0.471, 0.495)$\\

 &  & 1.0 & 2.72 & $8.49$ & $-5.93$ & 0.121 & $(-0.722, 0.735)$ & $(-0.719, 0.733)$ & $(-0.753, 0.763)$\\

\multirow{-18}{*}{\centering\arraybackslash I} & \multirow{-6}{*}{\centering\arraybackslash $\tau_{2, 3}$} & 2.0 & 7.39 & $4.67$ & $-1.94$ & 0.085 & $(-0.897, 0.898)$ & $(-0.897, 0.899)$ & $(-0.913, 0.914)$\\
\cmidrule{1-10}
 &  & 0.0 & 1.00 & $-13.40$ & $-13.40$ & 0.188 & $(-0.047, -0.047)$ & $(-0.084, -0.084)$ & $(-0.261, 0.097)$\\

 &  & 0.1 & 1.11 & $-10.14$ & $-16.80$ & 0.197 & $(-0.133, 0.040)$ & $(-0.166, 0.000)$ & $(-0.332, 0.184)$\\

 &  & 0.2 & 1.22 & $-7.29$ & $-20.24$ & 0.204 & $(-0.214, 0.124)$ & $(-0.242, 0.080)$ & $(-0.397, 0.263)$\\

 &  & 0.5 & 1.65 & $0.97$ & $-29.69$ & 0.239 & $(-0.427, 0.352)$ & $(-0.442, 0.300)$ & $(-0.561, 0.475)$\\

 &  & 1.0 & 2.72 & $9.99$ & $-42.90$ & 0.287 & $(-0.673, 0.638)$ & $(-0.676, 0.591)$ & $(-0.747, 0.713)$\\

 & \multirow{-6}{*}{\centering\arraybackslash $\tau_{1, 2}$} & 2.0 & 7.39 & $16.52$ & $-54.85$ & 0.393 & $(-0.891, 0.893)$ & $(-0.888, 0.872)$ & $(-0.912, 0.911)$\\
\cmidrule{2-10}
 &  & 0.0 & 1.00 & $-14.93$ & $-14.93$ & 0.205 & $(-0.031, -0.031)$ & $(-0.072, -0.072)$ & $(-0.244, 0.102)$\\

 &  & 0.1 & 1.11 & $-12.24$ & $-17.45$ & 0.207 & $(-0.117, 0.052)$ & $(-0.155, 0.007)$ & $(-0.314, 0.186)$\\

 &  & 0.2 & 1.22 & $-8.85$ & $-20.76$ & 0.219 & $(-0.201, 0.132)$ & $(-0.234, 0.083)$ & $(-0.379, 0.265)$\\

 &  & 0.5 & 1.65 & $-2.15$ & $-29.86$ & 0.240 & $(-0.422, 0.349)$ & $(-0.443, 0.295)$ & $(-0.556, 0.463)$\\

 &  & 1.0 & 2.72 & $7.20$ & $-42.13$ & 0.269 & $(-0.683, 0.626)$ & $(-0.689, 0.578)$ & $(-0.755, 0.700)$\\

 & \multirow{-6}{*}{\centering\arraybackslash $\tau_{1, 3}$} & 2.0 & 7.39 & $14.25$ & $-54.48$ & 0.371 & $(-0.903, 0.885)$ & $(-0.902, 0.863)$ & $(-0.921, 0.902)$\\
\cmidrule{2-10}
 &  & 0.0 & 1.00 & $-1.32$ & $-1.32$ & 0.128 & $(0.015, 0.015)$ & $(0.014, 0.014)$ & $(-0.108, 0.134)$\\

 &  & 0.1 & 1.11 & $-1.31$ & $-2.93$ & 0.122 & $(-0.072, 0.100)$ & $(-0.074, 0.098)$ & $(-0.195, 0.218)$\\

 &  & 0.2 & 1.22 & $0.49$ & $-3.65$ & 0.131 & $(-0.156, 0.177)$ & $(-0.158, 0.176)$ & $(-0.274, 0.290)$\\

 &  & 0.5 & 1.65 & $4.48$ & $-8.21$ & 0.136 & $(-0.378, 0.379)$ & $(-0.380, 0.375)$ & $(-0.476, 0.472)$\\

 &  & 1.0 & 2.72 & $10.47$ & $-14.50$ & 0.171 & $(-0.643, 0.622)$ & $(-0.642, 0.616)$ & $(-0.702, 0.679)$\\

\multirow{-18}{*}{\centering\arraybackslash II} & \multirow{-6}{*}{\centering\arraybackslash $\tau_{2, 3}$} & 2.0 & 7.39 & $18.63$ & $-20.99$ & 0.225 & $(-0.882, 0.862)$ & $(-0.879, 0.857)$ & $(-0.899, 0.881)$\\
\bottomrule
\end{tabular}
\begin{tablenotes}
\item[\dag] The first four columns represent the simulation settings: Scenario and ATE are the simulation scenarios and pairwise ATEs of interest, respectively; $\lambda$ is the sensitivity parameter and $\Lambda = exp(\lambda)$. The next five columns are the results: respectively percentage average bias (in SD units) in the lower and upper bound of the point estimate interval; the non-coverage rate of the confidence intervals under the proposed framework (the desired non-coverage rate is $10\%$); the partially identified point estimate intervals of the pairwise ATEs; the median interval for the point estimates; the median confidence interval.
\end{tablenotes}

\end{threeparttable}}
\end{table}

We simulate $1000$ datasets for each scenario and estimate the interval of point estimates and construct $90\%$ confidence intervals for the pairwise ATEs under the proposed sensitivity analysis for multivalued treatment settings. We report the percentage average bias (in SD units) in the lower and upper bounds of the point estimate interval, the non-coverage rate of the confidence interval, the median interval of point estimates, and the median confidence intervals calculated from the $1000$ simulated datasets.

The simulation results under the proposed framework of sensitivity analysis are presented in Table \ref{tab:multivalued-RR}. When there is adequate overlap (Scenario-I), we observe that the average bias in the SIPW point estimators of the pairwise ATEs lies within $10\%$ of its standard deviation in almost all cases under under the proposed framework. The percentile bootstrap confidence intervals also satisfy the nominal $90\%$ coverage rate in the presence of adequate overlap. However, we observe that the performance of the SIPW point estimators and the percentile bootstrap confidence intervals deteriorates when there is a lack of overlap (Scenario-II). These findings are similar to the ones observed by \cite{zhao2019sensitivity} and \cite{basit2023sensitivity}, and therefore, we recommend to interpret the sensitivity analysis results under the proposed framework with caution when there is a lack of overlap in the covariate distribution among different treatment groups.


\section{Real Data Application}
\label{sec:application}

In this section, we applied our proposed sensitivity analysis framework for multivalued treatments to estimate the causal effect of maternal education on the fertility rate in Bangladesh. The dataset used in this analysis was obtained from the latest round of the Bangladesh Multiple Indicator Cluster Survey (MICS) 2019 \citep{mics2019bangladesh}. We considered the number of children ever born to women in the reproductive age ($15-49$ years) as a measure of fertility, which is the outcome of interest. The treatment variable was the maternal education level that has four levels- "pre-primary or none", "primary", "secondary", and "higher secondary or beyond". The following variables were considered as the observed confounders, which were assumed to be associated with both treatment and outcome:  place of residence ("urban" or "rural"), district, religion ("muslim" or "non-muslim"), household's wealth index, women's age at marriage, and education level of the household head \citep{tomal2022weighted}. After discarding missing values from the outcome, treatment, and the observed confounders, we obtained complete data on  $53,481$ women aged $15-49$ years to conduct the analysis.

Using the proposed sensitivity analysis framework, we conduct the sensitivity analysis for the three pairwise ATEs between "pre-primary or none", "primary", and "secondary" vs. "higher secondary or beyond" level of maternal education, denoted by  $\tau_{1, 4}$, $\tau_{2, 4}$, and $\tau_{3, 4}$, respectively. We considered $\Gamma_0 = \exp(\gamma_0) = \{1, 1.25, 1.5, 1.75, 2.00, 2.25, 2.50, 2.75\}$ to assess the sensitivity of the causal effect estimates to the presence of varying magnitudes of unmeasured confounding. As discussed in Remark \ref{remark: sens-parameter}, we did not need to specify the values of the sensitivity parameter $\Gamma_1$ to conduct the sensitivity analysis. Since the treatment variable is ordinal, we fitted an ordinal logistic regression model, namely the continuation ratio regression model, to estimate the generalized propensity scores (GPSs) using the chosen observed confounders. The estimated GPSs ranged from $0.01$ to $0.98$ with a mean of $0.25$ across the four treatment groups. In light of our findings in the simulation studies in Section \ref{sec:multivalued_sim}, we recommend to interpret the results obtained from the SIPW estimators with caution as some of the estimated GPSs are close to zero.

\begin{table}[!h]

\caption{SIPW point estimate intervals and 90\% percentile bootstrap confidence intervals for the pairwise ATEs under the RR framework for different values of the sensitivity parameter $\Gamma_0$. The pariwise ATEs of pre-primary or no education, primary education, and secondary education vs. higher seondary or beyond education are denoted by  $\tau_{1, 4}$, $\tau_{2, 4}$, and $\tau_{3, 4}$, respectively}
\label{tab:mics-sensitivity}
\centering
\begin{tabular}[t]{crrrr}
\toprule
Estimand & $\Gamma_0$ & $\gamma_0$ & \makecell[c]{Point estimate \\ interval} & \makecell[c]{90\% confidence \\ Interval}\\
\midrule
 & 1 & 0.00 & $(1.97, 1.97)$ & $(1.92, 2.02)$\\

 & 1.25 & 0.22 & $(1.59, 2.46)$ & $(1.52, 2.53)$\\

 & 1.5 & 0.41 & $(1.22, 2.83)$ & $(1.14, 2.90)$\\

 & 1.75 & 0.56 & $(0.91, 3.14)$ & $(0.83, 3.21)$\\

 & 2 & 0.69 & $(0.66, 3.39)$ & $(0.59, 3.46)$\\

 & 2.25 & 0.81 & $(0.46, 3.61)$ & $(0.38, 3.68)$\\

 & 2.5 & 0.92 & $(0.29, 3.82)$ & $(0.22, 3.90)$\\

\multirow{-8}{*}{\centering\arraybackslash $\tau_{1, 4}$} & 2.75 & 1.01 & $(0.13, 4.01)$ & $(0.03, 4.09)$\\
\cmidrule{1-5}
 & 1 & 0.00 & $(1.46, 1.46)$ & $(1.43, 1.48)$\\

 & 1.25 & 0.22 & $(1.15, 1.95)$ & $(1.10, 1.99)$\\

 & 1.5 & 0.41 & $(0.82, 2.27)$ & $(0.77, 2.32)$\\

 & 1.75 & 0.56 & $(0.56, 2.57)$ & $(0.50, 2.63)$\\

 & 2 & 0.69 & $(0.30, 2.83)$ & $(0.24, 2.88)$\\

 & 2.25 & 0.81 & $(0.07, 3.06)$ & $(0.02, 3.11)$\\

 & 2.5 & 0.92 & $(-0.12, 3.24)$ & $(-0.17, 3.28)$\\

\multirow{-8}{*}{\centering\arraybackslash $\tau_{2, 4}$} & 2.75 & 1.01 & $(-0.28, 3.39)$ & $(-0.33, 3.44)$\\
\cmidrule{1-5}
 & 1 & 0.00 & $(0.75, 0.75)$ & $(0.72, 0.80)$\\

 & 1.25 & 0.22 & $(0.45, 1.17)$ & $(0.39, 1.23)$\\

 & 1.5 & 0.41 & $(0.13, 1.49)$ & $(0.08, 1.54)$\\

 & 1.75 & 0.56 & $(-0.13, 1.75)$ & $(-0.18, 1.80)$\\

 & 2 & 0.69 & $(-0.33, 1.97)$ & $(-0.38, 2.01)$\\

 & 2.25 & 0.81 & $(-0.49, 2.13)$ & $(-0.54, 2.17)$\\

 & 2.5 & 0.92 & $(-0.62, 2.27)$ & $(-0.67, 2.31)$\\

\multirow{-8}{*}{\centering\arraybackslash $\tau_{3, 4}$} & 2.75 & 1.01 & $(-0.73, 2.38)$ & $(-0.78, 2.41)$\\
\bottomrule
\end{tabular}
\end{table}

\begin{figure}[t!]
\centering
\includegraphics{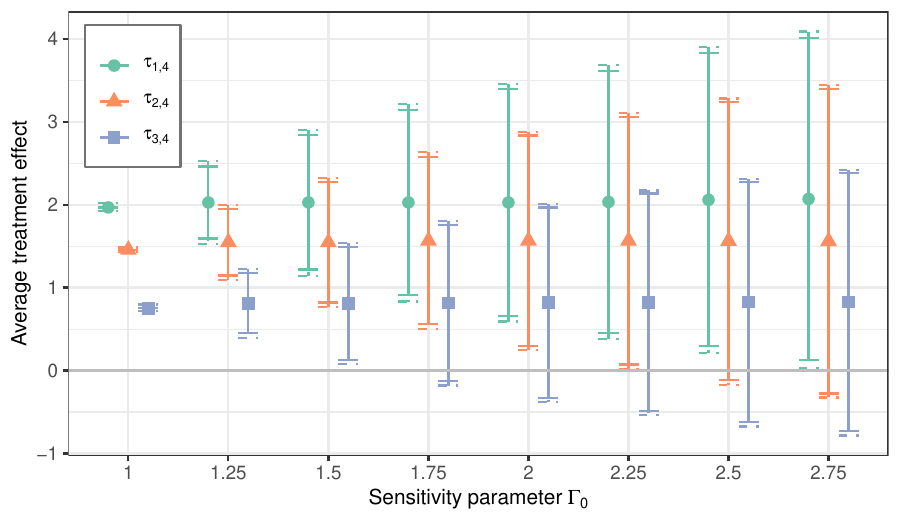}
\caption{Graphical representation of the sensitivity analysis results for the three pairwise ATEs under the RR framework. The solid error bars are the range of point estimates and the dashed error bars (together with the solid bars) are the confidence intervals. The circles/triangles/squares are the mid-points of the solid bars.}
\label{fig:sensitivity-plot}
\end{figure}

For each value of the sensitivity parameter $\Gamma_0$, we estimated the partially identified point estimate intervals and obtained $100(1 - \alpha)\%$ confidence intervals using percentile bootstrap with $B=1000$. Table \ref{tab:mics-sensitivity} and Figure \ref{fig:sensitivity-plot} represent the sensitivity analysis results under the risk-ratio-based framework. We can observe that in the case of the pairwise ATE between "pre-primary or none" and "higher secondary or beyond" level of education ($\tau_{1, 4}$), the confidence intervals do not contain the null value of zero, suggesting a significant non-zero causal effect, for at least $\Gamma_0 = 2.75$. That is, the estimated pairwise ATE $\tau_{1,4}$ is significantly different from zero even in the presence of unmeasured confounders that cause the true (unobserved) GPSs to be 2.75 times higher than the observed GPSs. Similarly, the ATE between the "primary" and "higher secondary or beyond" level of education ($\tau_{2,4}$) is statistically significant for at least $\Gamma_0=2.25$. However, the ATE between "secondary" and "higher secondary or beyond" education ($\tau_{3,4}$) is insignificant for values of $\Gamma_0$ as small as about $1.5$.

Based on the conducted sensitivity analysis under our proposed framework, we can imply that the estimated pairwise ATE between "pre-primary or none" vs "higher secondary or beyond" education (\(\tau_{1,4}\)) and "primary" vs "higher secondary or beyond" education (\(\tau_{2,4}\)) are less sensitive to the presence of unmeasured confounding in that very strong unmeasured confounders are needed to invalid the causal estimates of \(\tau_{1,4}\) and \(\tau_{2,4}\). This can be perceived as a substantial evidence of significant causal effect of maternal education on the fertility of women in Bangladesh.

\section{Conclusion}
\label{sec:conclusion}

Sensitivity analysis is crucial for the accurate interpretation of causal conclusions drawn from observational studies. In this paper, we proposed a risk-ratio-based modified marginal sensitivity model by extending the odds-ratio-based sensitivity model of \citet{zhao2019sensitivity}. We further extended our proposed sensitivity model to observational studies with multivalued treatments. As the sensitivity parameter in our proposed framework quantifies the degree of unmeasured confounding using risk ratios of the observed and true (generalized) propensity scores, it is easier and more intuitive to interpret the sensitivity analysis results under the proposed framework compared to that under odds-ratio-based frameworks. We illustrated that estimation of the inverse probability weighting (IPW) causal effect estimators under our proposed framework can be computed efficiently using a percentile bootstrap approach. We also conducted simulation studies that suggest that our proposed sensitivity analysis framework performs well when there is an adequate overlap among the treatment groups. However, the performance deteriorates due to the instability of the inverse probability weights when there is a lack of overlap. Finally, we demonstrated sensitivity analysis under the proposed framework using an empirical study, where we have estimated the causal effect of maternal education on the female fertility in Bangladesh.

There are a number of further potential research directions. We are currently working on incorporating other smooth causal effect estimators, such as the doubly-robust augmented IPW (AIPW) estimators \citep{robins1994estimation} and generalized overlap weighting (GOW) estimators \citep{li2019propensity} into our proposed framework. Furthermore, we intend to work on calibrating the the sensitivity parameter in our framework to the observed confounders. \citet{zhang2020calibrated} have recently worked on such calibration of the sensitivity model of \citet{gastwirth1998dual} in matched observational studies.

\bibliographystyle{biometrika}
\bibliography{manuscript_v2}

\begin{thebibliography}{24}
\expandafter\ifx\csname natexlab\endcsname\relax\def\natexlab#1{#1}\fi

\bibitem[{Basit et~al.(2023)Basit, Latif \& Wahed}]{basit2023sensitivity}
\textsc{Basit, M.~A.}, \textsc{Latif, M.~A.} \& \textsc{Wahed, A.~S.} (2023).
\newblock Sensitivity analysis for causal effects in observational studies with
  multivalued treatments.
\newblock \textit{arXiv preprint arXiv:2308.15986} .

\bibitem[{BBS \& UNICEF(2019)}]{mics2019bangladesh}
\textsc{BBS} \& \textsc{UNICEF} (2019).
\newblock Bangladesh multiple indicator cluster survey 2019, progotir pathey.

\bibitem[{Bind \& Rubin(2021)}]{bind2021importance}
\textsc{Bind, M.-A.~C.} \& \textsc{Rubin, D.} (2021).
\newblock The importance of having a conceptual stage when reporting
  non-randomized studies.
\newblock \textit{Biostatistics \& Epidemiology} \textbf{5}, 9--18.

\bibitem[{Bonvini et~al.(2022)Bonvini, Kennedy, Ventura \&
  Wasserman}]{bonvini2022sensitivity}
\textsc{Bonvini, M.}, \textsc{Kennedy, E.}, \textsc{Ventura, V.} \&
  \textsc{Wasserman, L.} (2022).
\newblock Sensitivity analysis for marginal structural models.
\newblock \textit{arXiv preprint arXiv:2210.04681} .

\bibitem[{Carnegie et~al.(2016)Carnegie, Harada \&
  Hill}]{carnegie2016assessing}
\textsc{Carnegie, N.~B.}, \textsc{Harada, M.} \& \textsc{Hill, J.~L.} (2016).
\newblock Assessing sensitivity to unmeasured confounding using a simulated
  potential confounder.
\newblock \textit{Journal of Research on Educational Effectiveness} \textbf{9},
  395--420.

\bibitem[{Charnes \& Cooper(1962)}]{charnes1962programming}
\textsc{Charnes, A.} \& \textsc{Cooper, W.~W.} (1962).
\newblock Programming with linear fractional functionals.
\newblock \textit{Naval Research Logistics Quarterly} \textbf{9}, 181--186.

\bibitem[{Cornfield et~al.(1959)Cornfield, Haenszel, Hammond, Lilienfeld,
  Shimkin \& Wynder}]{cornfield1959smoking}
\textsc{Cornfield, J.}, \textsc{Haenszel, W.}, \textsc{Hammond, E.~C.},
  \textsc{Lilienfeld, A.~M.}, \textsc{Shimkin, M.~B.} \& \textsc{Wynder, E.~L.}
  (1959).
\newblock Smoking and lung cancer: recent evidence and a discussion of some
  questions.
\newblock \textit{Journal of the National Cancer Institute} \textbf{22},
  173--203.

\bibitem[{Dorn et~al.(2021)Dorn, Guo \& Kallus}]{dorn2021doubly}
\textsc{Dorn, J.}, \textsc{Guo, K.} \& \textsc{Kallus, N.} (2021).
\newblock Doubly-valid/doubly-sharp sensitivity analysis for causal inference
  with unmeasured confounding.
\newblock \textit{arXiv preprint arXiv:2112.11449} .

\bibitem[{Gastwirth et~al.(1998)Gastwirth, Krieger \&
  Rosenbaum}]{gastwirth1998dual}
\textsc{Gastwirth, J.~L.}, \textsc{Krieger, A.~M.} \& \textsc{Rosenbaum, P.~R.}
  (1998).
\newblock Dual and simultaneous sensitivity analysis for matched pairs.
\newblock \textit{Biometrika} \textbf{85}, 907--920.

\bibitem[{Imbens(2000)}]{imbens2000role}
\textsc{Imbens, G.~W.} (2000).
\newblock The role of the propensity score in estimating dose-response
  functions.
\newblock \textit{Biometrika} \textbf{87}, 706--710.

\bibitem[{Kallus et~al.(2019)Kallus, Mao \& Zhou}]{kallus2019interval}
\textsc{Kallus, N.}, \textsc{Mao, X.} \& \textsc{Zhou, A.} (2019).
\newblock Interval estimation of individual-level causal effects under
  unobserved confounding.
\newblock In \textit{The 22nd international conference on artificial
  intelligence and statistics}. PMLR.

\bibitem[{Kang et~al.(2007)Kang, Schafer et~al.}]{kang2007demystifying}
\textsc{Kang, J.~D.}, \textsc{Schafer, J.~L.} et~al. (2007).
\newblock Demystifying double robustness: A comparison of alternative
  strategies for estimating a population mean from incomplete data.
\newblock \textit{Statistical Science} \textbf{22}, 523--539.

\bibitem[{Li et~al.(2019)}]{li2019propensity}
\textsc{Li, F.} et~al. (2019).
\newblock Propensity score weighting for causal inference with multiple
  treatments.
\newblock \textit{The Annals of Applied Statistics} \textbf{13}, 2389--2415.

\bibitem[{Robins(2002)}]{robins2002covariance}
\textsc{Robins, J.~M.} (2002).
\newblock Comment on `{Covariance} adjustment in randomized experiments and
  observational studies'.
\newblock \textit{Statistical Science} \textbf{17}, 309--321.

\bibitem[{Robins et~al.(1994)Robins, Rotnitzky \& Zhao}]{robins1994estimation}
\textsc{Robins, J.~M.}, \textsc{Rotnitzky, A.} \& \textsc{Zhao, L.~P.} (1994).
\newblock Estimation of regression coefficients when some regressors are not
  always observed.
\newblock \textit{Journal of the American Statistical Association} \textbf{89},
  846--866.

\bibitem[{Rosenbaum(1987)}]{rosenbaum1987sensitivity}
\textsc{Rosenbaum, P.~R.} (1987).
\newblock Sensitivity analysis for certain permutation inferences in matched
  observational studies.
\newblock \textit{Biometrika} \textbf{74}, 13--26.

\bibitem[{Rosenbaum(2002)}]{rosenbaum2002b}
\textsc{Rosenbaum, P.~R.} (2002).
\newblock \textit{Observational studies}.
\newblock Springer.

\bibitem[{Rosenbaum et~al.(2002)}]{rosenbaum2002covariance}
\textsc{Rosenbaum, P.~R.} et~al. (2002).
\newblock Covariance adjustment in randomized experiments and observational
  studies.
\newblock \textit{Statistical Science} \textbf{17}, 286--327.

\bibitem[{Rubin(1978)}]{rubin1978bayesian}
\textsc{Rubin, D.~B.} (1978).
\newblock Bayesian inference for causal effects: The role of randomization.
\newblock \textit{The Annals of Statistics} \textbf{6}, 34--58.

\bibitem[{Tomal et~al.(2022)Tomal, Khan \& Wahed}]{tomal2022weighted}
\textsc{Tomal, J.~H.}, \textsc{Khan, J.~R.} \& \textsc{Wahed, A.~S.} (2022).
\newblock Weighted bayesian poisson regression for the number of children ever
  born per woman in bangladesh.
\newblock \textit{Journal of Statistical Theory and Applications} \textbf{21},
  79--105.

\bibitem[{VanderWeele \& Ding(2017)}]{vanderweele2017sensitivity}
\textsc{VanderWeele, T.~J.} \& \textsc{Ding, P.} (2017).
\newblock Sensitivity analysis in observational research: introducing the
  e-value.
\newblock \textit{Annals of internal medicine} \textbf{167}, 268--274.

\bibitem[{Yadlowsky et~al.(2022)Yadlowsky, Namkoong, Basu, Duchi \&
  Tian}]{yadlowsky2022bounds}
\textsc{Yadlowsky, S.}, \textsc{Namkoong, H.}, \textsc{Basu, S.},
  \textsc{Duchi, J.} \& \textsc{Tian, L.} (2022).
\newblock Bounds on the conditional and average treatment effect with
  unobserved confounding factors.
\newblock \textit{The Annals of Statistics} \textbf{50}, 2587--2615.

\bibitem[{Zhang \& Small(2020)}]{zhang2020calibrated}
\textsc{Zhang, B.} \& \textsc{Small, D.~S.} (2020).
\newblock A calibrated sensitivity analysis for matched observational studies
  with application to the effect of second-hand smoke exposure on blood lead
  levels in children.
\newblock \textit{Journal of the Royal Statistical Society Series C: Applied
  Statistics} \textbf{69}, 1285--1305.

\bibitem[{Zhao et~al.(2019)Zhao, Small \& Bhattacharya}]{zhao2019sensitivity}
\textsc{Zhao, Q.}, \textsc{Small, D.~S.} \& \textsc{Bhattacharya, B.~B.}
  (2019).
\newblock Sensitivity analysis for inverse probability weighting estimators via
  the percentile bootstrap.
\newblock \textit{Journal of the Royal Statistical Society: Series B
  (Statistical Methodology)} \textbf{81}, 735--761.

\end{thebibliography}

\end{document}